# Comparison between illumination model and hydrodynamic simulation for a Direct Drive laser irradiated target

M. TEMPORAL[1,a], B. CANAUD[2], W. J. GARBETT[3], and R. RAMIS[4]

[1]*Centre de Mathématiques et de Leurs Applications, ENS Cachan and CNRS, 61 Av. du President Wilson, F-94235 Cachan Cedex, France*

[2]*CEA, DIF, F-91297, Arpajon Cedex, France*

[3]*AWE plc, Aldermaston, Reading, Berkshire, RG7 4PR, United Kingdom*

[4]*ETSI Aeronáuticos, Universidad Politécnica de Madrid, 28040 Madrid, Spain*

**Abstract**
A spherical target irradiated by laser beams located at 49º and 131º with respect to the polar axis has been considered. The illumination model has been used to evaluate the irradiation non-uniformity assuming circular and elliptical super-gaussian laser intensity profiles and the irradiation scheme has been optimized by means of the Polar Direct Drive technique. A parametric study has been performed providing the irradiation non-uniformity as a function of the Polar Direct Drive displacement and of the laser intensity profile parameters. Moreover, two-dimensional axis-symmetric hydrodynamic simulations have been performed for a plastic sphere irradiated by laser beams characterized by a constant flat temporal power pulse. In these simulations the front of the inward shock wave has been tracked providing the time-evolution of any non-uniformity. The results provided by the two methods - illumination model and hydrodynamic data - have been compared and it is found that the illumination model reproduces the main behaviour exhibited by the hydrodynamic data. The two models provide compatible values for the optimum Polar Direct Drive parameter and similar optimal super-gaussian profiles.

[a] mauro.temporal@hotmail.com

## 1. INTRODUCTION

In Inertial Confinement Fusion (ICF) (Lindl, 1995, 2004; Atzeni & Meyer-ter-Vehn, 2004) a spherical target containing the Deuterium-Tritium thermonuclear fuel is heated and compressed to generate the ignition of the nuclear fusion reactions. In the central ignition scheme the ignition conditions are generated in a relatively small plasma volume characterized by a plasma temperature around $T \approx 10$ keV and a confinement parameter $\rho R$ of about 0.3 g/cm$^2$ ($\alpha$-particle range).

In the Indirect Drive (Lindl, 1995) approach the fusion capsule is located within a high-Z casing. Powerful laser beams are directed into the case, where a fraction of energy is converted to x-rays, driving the capsule implosion. This scheme has been adopted by the National Ignition Facility (NIF) (Miller *et al.*, 2004; Moses *et al.*, 2009; Lindl *et al.*, 2014) (Livermore, USA) and by the Laser MegaJoule (LMJ) (Cavailler, 2005; Lion, 2010) facility (Bordeaux, France). The NIF facility is composed by 192 laser beams arranged in 48 quads and has demonstrated a total energy (power) of 2 MJ (500 TW), whilst the current design of the LMJ consist of 176 laser beams for a total of 44 quads (1.3 MJ, 440 TW).

Another option is offered by the Direct Drive scheme (Nuckolls *et al.,* 1972; Bodner *et al.*, 2002) where the laser beams heat directly the external shell of the ICF capsule. In both cases the uniformity of the irradiation represents an important issue. Indeed, large non-uniformity in the irradiation would introduce inefficiency through asymmetric fuel assembly and could trigger dangerous hydrodynamic instabilities as Richtmyer-Meshkov and Rayleigh-Taylor. These instabilities can cause deleterious mixing of shell material into the fuel or could damage and even destroy the capsule during the implosion.

The illumination model (Skupsky & Lee, 1983; Schmitt, 1984) provides a simple way to evaluate the non-uniformity of the irradiation for a given laser-capsule configuration. The model can also include statistical analyses that take into account beam uncertainties such as power-imbalance, pointing error and target positioning. Several studies have been performed to analyze the capsule illumination uniformity for different facilities (Murakami, *et al.*, 1993, 1995, 2010; Canaud *et al.*, 2002; Temporal *et al.*, 2009, 2010a, 2010b, 2011a, 2011b, 2014a, 2014b). In these studies it has been assumed that the quality of the illumination (usually measured by a root-mean-square deviation of the incident intensity on a spherical surface) is representative of the non-uniformity induced in the first shock wave, also called imprint phase.

The aim of this paper is to test if this assumption is satisfied for the case of a specific two-dimensional axis-symmetric irradiation configuration defined in Sec. II. Two-dimensional hydrodynamic calculations have been performed to analyze the non-uniformity of the shock front generated in a spherical plastic target (Sec. III). A parametric study, varying the laser beam intensity profile and the Polar Direct Drive (Skupsky *et al.*, 2004; Craxton *et al.*, 2005) parameter, has been performed using both the illumination model and the hydrodynamic model. Finally, the results of the hydrodynamic calculations have been compared with the data obtained with the illumination model (Sec IV). We anticipate that the results of the two models have shown a good agreement, enabling the use of the illumination model to define the optimum laser-capsule parameters that optimize the non-uniformity of the first shock wave.

## 2. IRRADIATION CONFIGURATION

The LMJ facility configuration considered in this article foresees the use of a total of 176 laser beams organized in 44 quads (3ω, 1.3 MJ, 440 TW). Four quads will be devoted to diagnostics and the other 40 quads are distributed in four cones – two per hemisphere – located at 33.2°, 146.8° (1$^{st}$ cone) and at 49°, 131° (2$^{nd}$ cone) with respect to the polar axis. The LMJ facility has been planned and optimized for the indirect drive scheme. Nevertheless, it could also be useful in order to test aspects relevant for the Direct Drive approach. In this context special attention is devoted to the Shock Ignition scheme (Betti *et al.*, 2007) that envisages the employ of two laser pulses: a first compression pulse followed by a high-power igniting pulse. One of the possibilities offered by the LMJ facility is to dedicate the 20 quads in the 2$^{nd}$ cone to the compression pulse and the 20 quads at the 1$^{st}$ cone for the ignition phase. Recently it has been shown (Canaud *et al.*, 2012) that special attention has to be paid to the sphericity of fuel assembly even when using shock ignition.

This paper aims to analyze the irradiation non-uniformity provided during the first few ns that dominate the imprinting phase. The first shock is important as it is principally responsible for determining the entropy of the fuel. Moreover, the asymmetry of the shock needs to be minimised and studies performed at NIF suggest that one needs to tune the symmetry of the first and fourth shocks (Landen *et al.*, 2011; Kyrala *et al.*, 2011). Thus we only considered the compression pulse provided by the 20 quads (10 per hemisphere) located in the 2$^{nd}$ cone. In a Direct Drive scheme the laser pulse power is formed by a relatively low-power foot pulse followed by the main pulse that drives the fuel compression (McKenty *et al.*, 2004; Canaud *et al.*, 2007; Brandon *et al.*, 2013). Hereafter our analysis is restricted to the irradiation non-uniformity provided by the first few ns of the low-power foot pulse that has been schematically represented by a flat constant 2 TW power pulse.

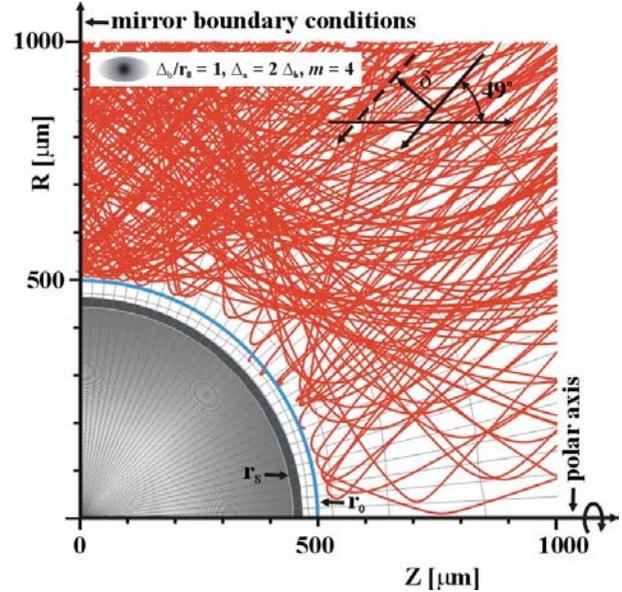

**Fig. 1.** Lagrangian cells at 2 ns, for a target whose initial radius is $r_0$ (blue curve). The red curves show some of the photon paths, while the white curve ($r_S$) is the position of the shock front.

The two-dimensional hydrodynamic calculations have been performed with the numerical code DUED (Atzeni, 1987). The code deals with tabulated EOS data, 2T model, flux limited heat conduction and inverse-bremsstrahlung laser energy deposition driven by a 3-dimensional ray-tracing package. In these hydrodynamic simulations a plastic (CH) spherical target with density $\rho_{CH} = 1.05$ g/cm$^3$ and radius $r_0 = 500$ μm has been considered. The system is axis-symmetric with respect to the polar axis (Z) and only a π/2 angular sector has been simulated assuming rigid boundary conditions at the plane of symmetry (see Fig. 1). The target has been discretized with a Lagrangian mesh (r-θ) composed by 32 cells equally distributed in the π/2 angular sector, while 300 cells are used in the radial dimension (50 cells equally-spaced between 0 to 200 μm and 250 cells distributed to have the same mass between r = 200 μm and $r_0$ = 500 μm).

The laser axis is located at 49° with respect to the polar axis and the intensity profile is given by the super-gaussian function $I(x,y) = I_0 \exp[-[(x/\Delta_a)^2+(y/\Delta_b)^2]^{m/2}$, characterized by the parameters $\Delta_a$ and $\Delta_b$ (half width at 1/e) and the exponent $m$. Of course, the elliptical laser intensity profile becomes circular when $\Delta_a = \Delta_b$. The y coordinate is located in the plane R-Z of the 2D hydro-calculations and it is orthogonal to the beam axis, while the $x$ coordinate is orthogonal to the R-Z plane. In agreement with the point design of the LMJ laser beams the ratio between the two axis of the elliptical profile is set to 2 ($\Delta_a$:$\Delta_b$ = 2:1) and the super-gaussian exponent is set to $m = 4$. At each hydrodynamic time step the 3D ray-tracing package follows the path of 2048 beamlets that are randomly distributed in the surface where the laser intensity is larger than $I_0/100$.

Advancements on the optimization of the Direct Drive capsule irradiation (Canaud *et al.*, 2004, 2007) have been recently offered by the Polar Direct Drive (PDD). In this case the laser beams axis are not aligned with the capsule centre but are moved by a quantity $\delta$ toward the capsule equator. In the Fig. 1 is shown the Lagrangian mesh at $t = 2$ ns for a spherical target irradiated by an elliptical laser intensity profile. The red curves are the 3D photon paths projected into the R-Z plane (only 1/10 of the total beamlets are shown). In the same figure is also indicated the position of the initial target radius ($r_0$) and the current shock front position ($r_S$).

## 3. ILLUMINATION MODEL AND HYDRODYNAMIC CALCULATIONS

The quality of the irradiation could be estimated as the root-mean-square deviation of the laser intensity $I(\theta,\varphi)$ that illuminates a spherical surface. In the case of a system axis-symmetric around the polar axis Z the intensity over the target surface $I(\theta)$ only depends on the polar angle $\theta$ and the rms non-uniformity $\sigma_{2D}$ is given by the equation 1.

$$\sigma_{2D} = \left\{ \frac{1}{2} \int_0^\pi [I(\theta) - <I>]^2 \sin(\theta) d\theta \right\}^{1/2} / <I> \quad (1)$$

where $<I>$ is the average intensity over the whole spherical surface. Two first calculations have been performed for a circular laser intensity profile characterized by $\Delta_b = \Delta_a = r_0 = 500$ μm and $m = 4$ and an elliptical profile where $\Delta_a = 2 \Delta_b$. In both cases the lasers axes are aligned with the target centre, thus the Polar Direct Drive parameter is set to zero ($\delta = 0$). The numerical results given by the illumination method show that the elliptical laser intensity profile provides a non-uniformity $\sigma_{2D} = 5.2$ % much better than the one found with the circular profile, $\sigma_{2D} = 11.9$ %. A parametric study performing a variation of the PDD parameter $\delta$ from 0 to $\delta/r_0 = 20$ % allows determination of the optimum PPD parameter that minimizes the non-uniformity. It has been found that for the circular (elliptical) laser intensity profile the optimum PDD parameter is $\delta/r_0 = 13.7$ % ($\delta/r_0 = 7.4$ %) and reduces the non-uniformity to $\sigma_{2D} = 1.1$ % ($\sigma_{2D} = 1.3$ %).

In the hydrodynamic calculations a plastic target is irradiated by the laser beams and the position of the shock front moving inward to the target centre is tracked as a function of time. The shock front position $r_S(\theta_i,t)$ is defined as the location of the plasma density growth up to the double of the initial plastic density ($\rho_S = 2 \rho_{CH}$). The position of the shock front is tracked at the 33 angles $\theta_i = \theta_{i-1} + \Delta\theta$, with $\theta_1 = 0$ and $\Delta\theta = \pi / 64$. Thus, with the $r_S(\theta_i,t)$ it is possible to calculate the non-uniformity $\sigma_S(t)$ (root-means-square deviation) associated to the shock front surface. In the Fig. 2 are shown the flow-chart of the average radius $\underline{r}(t)$ evaluated for two hydrodynamic simulations. In the first case (a) the target has been irradiated by a circular laser intensity profile ($\Delta_a = \Delta_b = r_0$, $m = 4$, $\delta/r_0 = 13.7$ %); whilst in the second simulation (b) the intensity profile is elliptical ($\Delta_a = 2 \Delta_b$, $\Delta_b = r_0$, $m = 4$, $\delta/r_0 = 7.4$ %). In the two frames of Fig. 2 the red dashed curves are the average position of the shock front $\underline{r}_S(t)$. The shock wave is faster in the case (a) rather than in the case (b); this is due to the different laser intensity profiles that generate a larger laser-target coupling in the circular case (a) with comparison to the elliptical case (b). Indeed, in the circular case the energy absorption is $\eta_a = 85$ % whilst it is reduced to $\eta_b = 61$ % in the elliptical case.

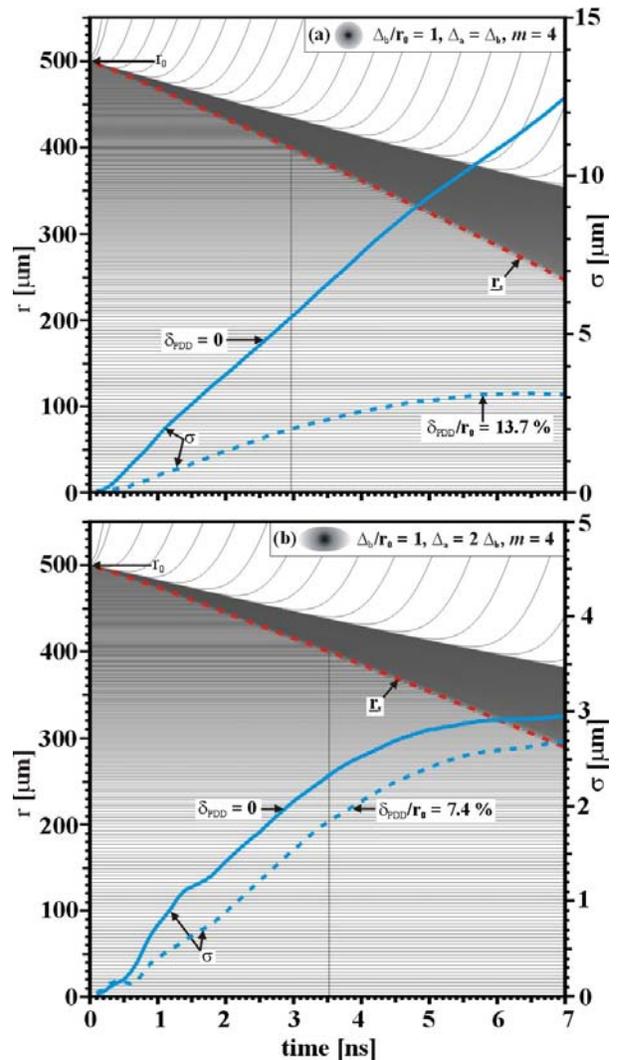

**Fig. 2.** Flow-chart of the average radial position $\underline{r}$ as a function of time for the circular (a) and elliptical (b) laser intensity profile. The blue curves represent the rms non-uniformity associated to the shock front without PDD (continuum) and with PDD (dashed). Red curves are the average shock front position, $\underline{r}_S$.

In both frames of Fig. 2 the blue curves are the rms non-uniformities $\sigma_S(t)$ associated to the shock front surface. The blue full curves refer to the cases without PDD whilst the dashed curves concern the cases with the PDD. It is found that the shock front non-uniformity increases with the time. Moreover, in the circular cases the PDD technique improves significantly the shock front uniformity. Differently, in the elliptical cases the application of the PDD technique modifies only slightly the uniformity. Indeed, when the shock wave arrive at the radius r = 400 μm (t ≈ 3 ns case (a) and t ≈ 3.5 ns case (b)) the circular laser intensity profile provides a rms non-uniformity $\sigma_c$ = 5.6 μm and the elliptical case $\sigma_e$ = 2.3 μm. Comparable non-uniformities are provided by the two intensity profiles when the optimum PDD parameters apply: $\sigma_c$ = 2.0 μm with $\delta_{PDD}/r_0$ = 13.7 % and $\sigma_e$ = 1.8 μm when $\delta_{PDD}/r_0$ = 7.4 %. Thus, in the circular (elliptical) case the PDD reduces the non-uniformity by a factor 2.8 (1.3).

## 4. PARAMETRIC ANALYSIS

In this section the non-uniformity of the target irradiation has been evaluated as a function of the PDD parameter $\delta$, the width of the circular and elliptical laser intensity profiles $\Delta_b$, and their super-gaussian exponent m. Then, these non-uniformities calculated with the illumination model and through the hydrodynamic calculation, have been compared.

As shown previously, in the hydrodynamic simulations the non-uniformity of the shock front evolves with the time. In order to be independent of the shock velocity - which depends on the laser intensity profile - the average non-uniformity $\sigma_n$ has been evaluated at ten positions of the shock front radius $r_n = r_0 - n\, 20$ μm, with n = [1-10]. These results are affected by numerical noise that produces a variation of about 10% in the value of the non-uniformity. In order to minimize this numerical noise the average non-uniformity $\sigma_n$ is evaluated by performing ten simulations.

In a first set of calculations with circular (elliptical) laser intensity profiles has been assumed a width $\Delta_b = r_0$ with $\Delta_a = \Delta_b$ ($\Delta_a = 2\,\Delta_b$) and m = 4. In these calculations the Polar Direct Drive technique is applied and the parameter $\delta$ varies between 0 to 100 μm (20% $r_0$). The shock front non-uniformities $\sigma_n$ (gray curves) as a function of $\delta$ are shown in the Fig. 3 for the circular and elliptical laser intensity profiles. The gray dashed curves enlighten the non-uniformity of the shock front at the selected radius of $r_5$ = 400 μm. The non-uniformities $\sigma_{2D}$ evaluated with the illumination model (blue curves) are also shown. It is found that the results provided by the two models exhibit similar behaviours. The illumination model provides an optimum PDD parameter $\delta_C/r_0$ = 13.7% and $\delta_E/r_0$ = 7.4% for the circular and elliptical cases, respectively. The hydrodynamic data show a minimum of the non-uniformity that corresponds quite well to these optimum PDD parameters. It is noted that in the elliptical case the optimum PDD parameter coincide with the illumination model at early time, whilst it shifts at lower values as the shock front moves deeper into the target.

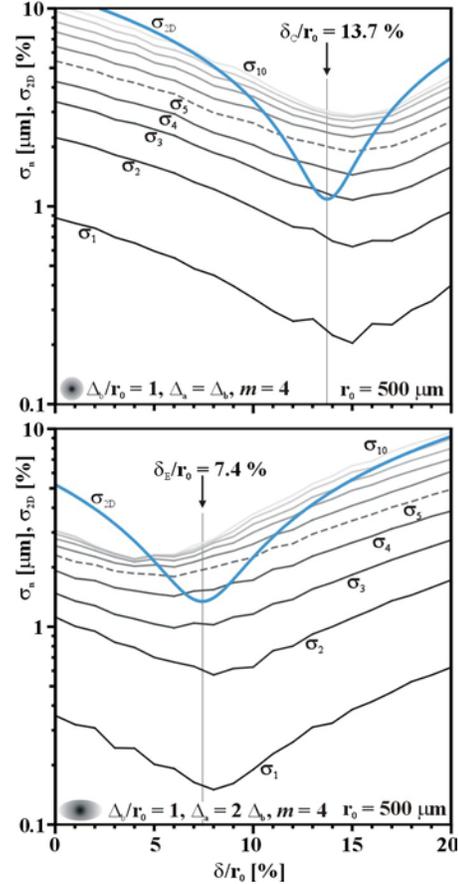

**Fig. 3.** Shock front non-uniformities $\sigma_n$ (gray curves) evaluated at different radius (between $r_1$ = 480 μm and $r_{10}$ = 300 μm, one each 20 μm) and illumination non-uniformity $\sigma_{2D}$ (blue curves) as a function of the PDD parameter $\delta/r_0$ for circular laser intensity profile (top) and elliptical case (bottom).

A second set of simulations has been performed keeping constant the optimum PDD parameters ($\delta_C/r_0$ = 13.7% and $\delta_E/r_0$ = 7.4%) and varying the width of the circular and elliptical laser intensity profiles, whilst the super-gaussian exponent is fixed to m = 4. In these calculations the width $\Delta_b$ varies between 0.6 $r_0$ and 1.8 $r_0$ and for the circular (elliptical) profile it is assumed $\Delta_a = \Delta_b$ ($\Delta_a = 2\,\Delta_b$). As before the non-uniformities of the hydrodynamic calculations are calculated when the shock wave crosses the ten spherical shell located between 480 μm and 300 μm (one each 20 μm). The non-uniformities $\sigma_n$ (gray curves) evaluated with the circular (top) and elliptical laser intensity profile (bottom) are shown in the Fig. 4 as a function of the parameter $\Delta_b$. The non-uniformities evaluated by the illumination model are indicated by blue curves. It is shown a good correspondence between the optimal laser intensity width ($\Delta_b \approx r_0$) obtained by the two methods. In the elliptical cases the correspondence is better at early time when the shock front is located at few ten μm into the target. Nevertheless, in both cases - circular and elliptical laser intensity profiles - the hydrodynamic data show a very well defined minimum of the shock front non-uniformity in correspondence to the optimal values found by the illumination model.

calculations assuming the optimal PDD parameter ($\delta_C/r_0 = 13.7\%$ and $\delta_E/r_0 = 7.4\%$). The red curves show the results of the illumination model without PDD, whilst the blue curves refer to the cases optimized with PDD. These results indicate that the illumination model reconstruct the same behaviours shown by the hydrodynamic data. In the elliptical case the curves of the non-uniformities with and without PDD cross each other. It is also worth noting that this behaviour is also shown by the hydrodynamic data. The hydrodynamic data appear smoothed and exhibit less pronounced minima in comparison with the results of the illumination model; nevertheless, both models show similar trends and similar optimum parameters that minimize the non-uniformity.

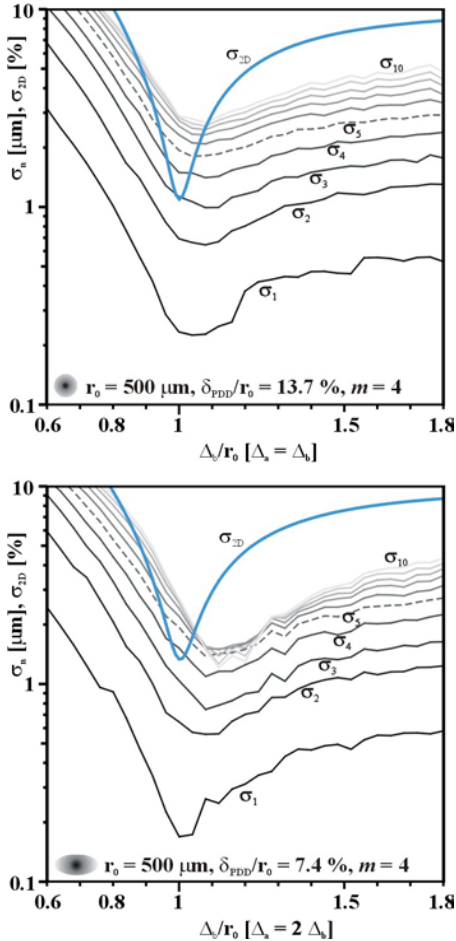

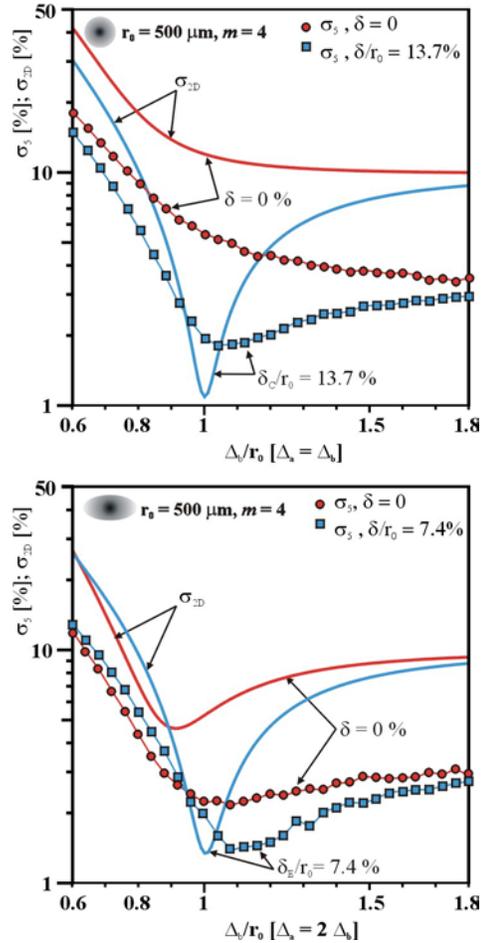

**Fig. 4.** Non-uniformities as a function of the width of laser intensity profiles, $\Delta_b$. Hydrodynamic data (gray curves) and illumination model (blue curves).

In the Fig. 5 is compared in detail the non-uniformity of the shock front when it crosses the surface located at the radius $r_5 = 400$ μm. These calculations assume a circular (top frame) and elliptical (bottom) laser intensity profile. In this case the shock front non-uniformity has been normalized to the distance travelled by the shock, namely $\sigma_5[\%] = \sigma_5[\mu m] / (r_0-r_5)$. These non-uniformities are shown as a function of the laser intensity profile parameter, $\Delta_b$. In Fig. 5 the red circles indicate the results without PDD, whilst the blue boxes refer to the

**Fig. 5.** Illumination non-uniformity ($\sigma_{2D}$) with (blue curves) and without (red curves) PDD is shown as a function of $\Delta_b$. The non-uniformity $\sigma_5$ evaluated by hydrodynamic simulations at the shock front position $r_5 = 400$ μm is shown by the red circles without PDD and by the blue boxes with PDD.

The optimum illumination non-uniformity is a function of the laser intensity profile parameters $\Delta_b$ and $m$. This has been also recently shown (Temporal *et al.*, 2013) for the Orion facility (Hopps *et al.*, 2013) that with 5 + 5 laser beams located at 50° and 130° has a configuration similar to the LMJ with the 10 + 10 quads at 49° and 131° with respect to the polar axis. Thus, to explore the non-uniformity in the Δ-m space for the two optimal PDD parameters: $\delta_C/r_0 = 13.7\%$ and $\delta_E/r_0 = 7.4\%$. The parametric study has been performed varying the width $\Delta_b/r_0$ from 0.6 to 1.8 and the super-gaussian exponent $m$ between 2 and 6. The results of the illumination non-uniformity as a function of $\Delta_b$ and $m$ are shown in the Fig. 6a (circular profile) and Fig. 6b (elliptical). Hydrodynamic calculations have been performed to analyse the non-uniformity $\sigma_5(\Delta_b/r_0, m)$ associated to the shock front when it crosses the radius $r_5 = 400$ μm. The contour levels of these non-uniformities (normalized to $r_0$-$r_5$) are shown in Fig. 6c for the case of circular laser intensity profiles and 6d for the elliptical cases. It is found that in the circular and elliptical cases both models provide similar results. The hydrodynamic data showed slightly better values of the minimum non-uniformities for the cases of elliptical profiles (≈ 1.4 %) compared to the circular ones (≈ 1.8 %). As it can be seen by comparing Fig. 6a with 6c and 6b with 6d, the minimum of the non-uniformity is roughly located in the same parametric space, even if the illumination model (shadowed areas) underestimates the optimal width (Δ) by about 10% and identifies a smaller parameter $m$. These results indicate that the optimum laser intensity profile provided by the illumination model could be used as a first estimation to minimize the non-uniformity associated to the front surface of the first shock.

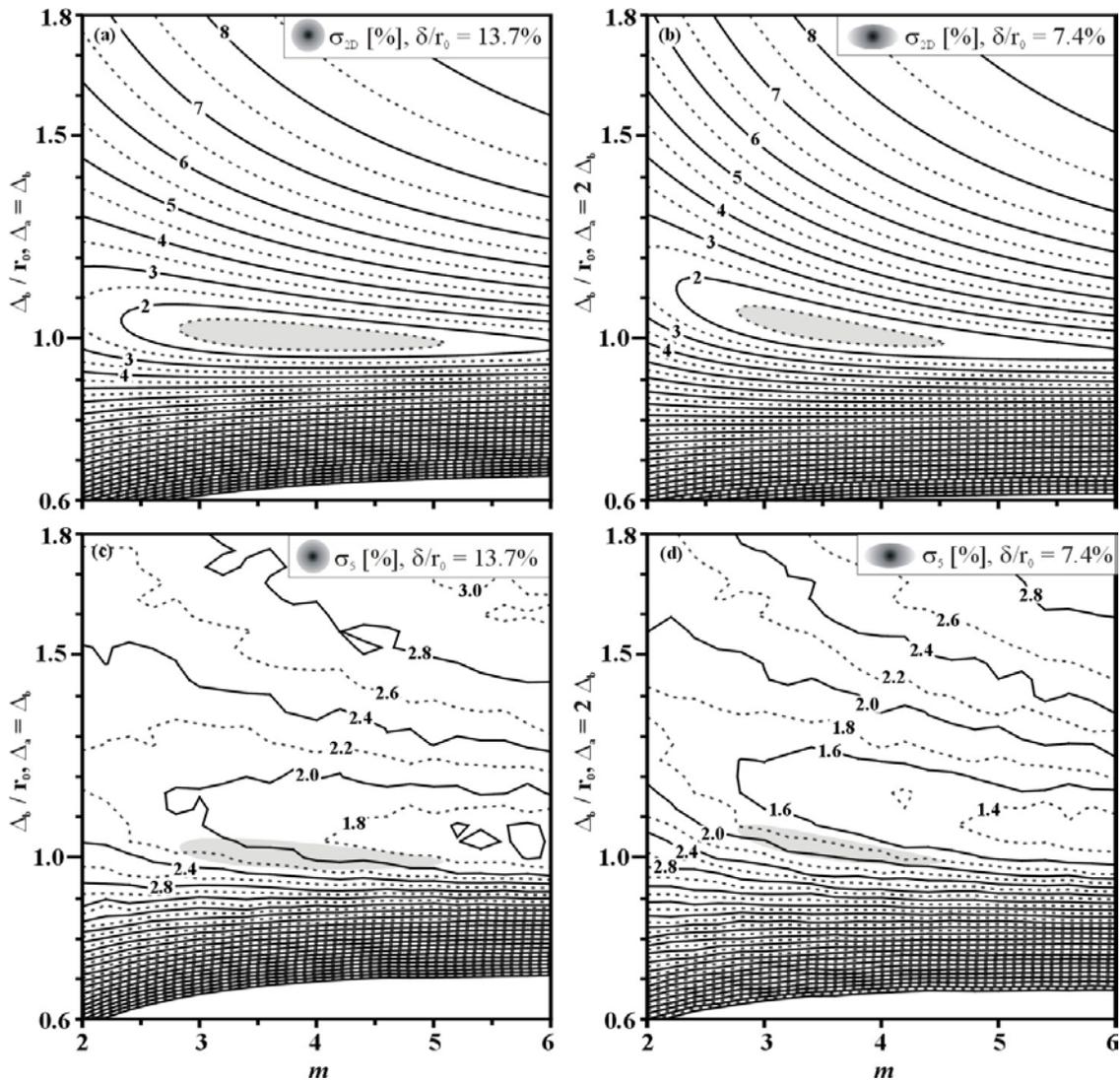

**Fig. 6.** Non-uniformities as a function of the laser intensity parameters $\Delta_b$ and $m$. Illumination model (a and b) and shock front non-uniformity $\sigma_5$ (c and d) provided by the hydrodynamic calculations at $r_5 = 400$ μm. Cases with circular profiles (a and c) assume a PDD parameter $\delta/r_0 = 13.7\%$ whilst in the elliptical cases (b and d) $\delta/r_0 = 7.4\%$. The shadowed areas correspond to the zones of minimum non-uniformity provided by the illumination model.

## 5. SUMMARY

A spherical target directly irradiated by laser beams has been considered. The laser configuration is axis-symmetric and the laser beams are located in two annular rings at the angles 49° and 131° with respect to the polar axis and correspond to those of the second ring in the Laser MegaJoule facility. The laser beams have been characterized by circular ($\Delta_a = \Delta_b$) or elliptical ($\Delta_a = 2\,\Delta_b$) super-gaussian laser intensity profile with the half-width at 1/e ($\Delta_a$ and $\Delta_b$) and the exponent $m$. In order to optimize the uniformity of the target irradiation the Polar Direct Drive technique has been used. In these cases the laser beams move by a quantity δ toward the target equatorial plane.

The non-uniformity associated to the laser irradiation has been calculated by means of the illumination model as well as using 2D hydrodynamic simulations. In these hydrodynamic calculations a spherical solid plastic (CH) target of radius $r_0 = 500$ μm has been considered. In order to mimic the foot-pulse of a Direct Drive irradiation a constant laser power of 2 TW has been associated to the super-gaussian laser intensity profile. A shock wave travelling inward through the target is generated as consequence of the laser irradiation. The shock front surface has been tracked providing a measurement of his non-uniformity during the time.

By using the illumination model it has been found that the minimum irradiation non-uniformity corresponds to the laser intensity parameters $\Delta_b/r_0 \approx 1$ and $3 < m < 5$ with the PDD parameters $\delta_C = 13.7\%$ and $\delta_E = 7.4\%$ for the circular and elliptical profile, respectively. The hydrodynamic results have been compared with those provided by the illumination model showing a satisfactory agreement between both approaches. In the hydrodynamic calculations the shock front non-uniformity grows with the time and the agreement of the two models in the optimal focal spot is better during the first few ns of the irradiation when the shock crosses the first 100 μm of the target.

In conclusion, the correlation between the illumination model and the non-uniformity of the first shock wave has been numerically confirmed for a specific two-dimensional axis-symmetric laser-target configuration. These results validate the hypothesis that the illumination model represents a valid method to assess the optimum laser-capsule parameters that minimise the shock front non-uniformity in the imprint phase. Nevertheless, further analysis is needed to explore the validity of this result also in the cases of more general laser-capsule configurations.

## ACKNOLEDGMENTS

M. T. and B. C. express their thanks to Daniel Bouche for the support given to this work. M. T. would like to thank Stefano Atzeni for the DUED code. R. R. was partially supported by the EURATOM/CIEMAT association in the framework of the "IFE Keep-in-Touch Activities".

HOPPS, N., DANSON, C., DUFFIELD, S., EGAN, D., ELSMERE, S., GIRLING, M., HARVEY, E., HILLIER, D., NORMAN, M., PARKER, S., TREADWELL, P., WINTER, D. & BETT, T. (2013). Overview of laser systems for the Orion facility at the AWE. *Appl. Opt.* **52**, 3597.

KYRALA, G.A., SEIFTER, A., KLINE, J.L., GOLDMAN, S.R., BATHA, S.H., & HOFFMAN, N.M. (2011). Tuning indirect-drive implosions using cone power balance. *Phys. Plasmas* **18**, 072703.

LANDEN, O.L., EDWARDS, J., HAAN, S.W., ROBEY, H.F., MILOVICH, J., SPEARS, B.K., WEBER, S.V., CLARK, D.S., LINDL, J.D., MACGOWAN, B.J., MOSES, E.I., ATHERTON, J., AMENDT, P.A., BOEHLY, T.R., BRADLEY, D.K., BRAUN, D.J., CALLAHAN, D.A., CELLIERS, P.M., COLLINS, J.W., DEWALD, E.L., DIVOL, L., FRENJE, J.A., GLENZER, S.H., HAMZA, A., HAMMEL, B.A., HICKS, D.G., HOFFMAN, N., IZUMI, N., JONES, O.S., KILKENNY, J.D., KIRKWOOD, R.K., KLINE, J.L., KYRALA, G.A., MARINAK, M.M., MEEZAN, N., MEYERHOFER, D.D., MICHEL, P., MUNRO, D.H., OLSON, R.E., NIKROO, A., REGAN, S.P., SUTER, L.J., THOMAS, C.A. & WILSON, D.C. (2011). Capsule implosion optimization during the indirect-drive National Ignition Campaign. *Phys. Plasmas* **18**, 051002.

LINDL, J. (1995). Development of the indirect-drive approach to inertial confinement fusion and the target physics basis for ignition and gain. *Phys. Plasmas* **2**, 3933–4024.

LINDL, J.D., AMENDT, P., BERGER, R.L., GLENDINNING, S.G., GLENZER, S.H., HAAN, S.W., KAUFFMAN, R.L., LANDEN, O.L. & SUTER L.J. (2004). The physics basis for ignition using indirect-drive targets on the National Ignition Facility. *Phys. Plasmas* **11**, 2, 339.

LINDL, J., LANDEN, O., EDWARDS, J., MOSES, E. & NIC TEAM (2014). Review of the National Ignition Campaign 2009-2012. *Phys. Plasmas* **21**, 020501.

LION, C. (2010). The LMJ program: an overview. *Journal of Physics: Conference Series* **244**, 012003.

McKENTY, P.W., SANGSTER, T.C., ALEXANDER, M., BETTI, R., CRAXTON, R.S., DELETREZ, J.A., ELASKY, L., EPSTEIN, R., FRANK, A., Yu. GLEBOV, V., GONCHAROV, V.N., HARDING, D.R., Jin, S., KNAUER, J.P., KECK, R.L., LOUCKS, S.J., LUND, L.D., McCRORY, R.L., MARSHALL, F.J., MEYERHOFER, D.D., REGAN, S.P., RADHA, P.B., ROBERTS, S., SEKA, W., SKUPSKY, S., SMALYUK, V.A., SOURES, J.M., THORP, K.A., WOZNIAK, M., FRENJE, J.A., LI, C.K., PETRASSO, R.D., SEGUIN, F.H., FLETCHER, K.A., PALADINO, S., FREEMAN, C., IZUMI, N., KOCH, J.A., LERCHE, R.A., MORAN, M.J., PHILLIPS, T.W., SCHMID, G.J. & SORCE, C. (2004). Direct-drive cryogenic target implosion performance on OMEGA. *Phys. Plasmas* **11**, 2790.

MILLER, G.H., MOSES, E.I. & WUEST, C.R. (2004). The National Ignition Facility: enabling fusion ignition for the 21st century. *Nucl. Fusion* **44**, S228.

MOSES, E.I., BOYD, R.N., REMINGTON, B.A., KEANE, C.J. & AL-AYAT, R. (2009). The National Ignition Facility: Ushering in a new age for high energy density science. *Phys. Plasmas* **16**, 041006.

MURAKAMI, M., NISHIHARA, K. & AZECHI, H. (1993). Irradiation non-uniformity due to imperfections of laser beams. *J. Appl.Phys.* **74**, 802–808.

MURAKAMI, M. (1995). Irradiation system based on dodecahedron for inertial confinement fusion. *Appl. Phys. Lett.* **66**, 1587.

MURAKAMI, M., SARUKURA, N., AZECHI, H., TEMPORAL, M. & SCHMITT, A.J. (2010). Optimization of irradiation configuration in laser fusion utilizing self-organizing electrodynamic system. *Phys. Plasmas* **17**, 082702.

NUCKOLLS, J., WOOD, L., THIESSEN, A. & ZIMMERMAN, G. (1972). Laser compression of matter to super-high densities: Thermonuclear (CTR) applications. *Nat.* **239**, 139–142.

SCHMITT, A.J. (1984). Absolutely uniform illumination of laser fusion pellets. *Appl. Phys. Lett.* **44**, 399-401.

SKUPSKY, S. & LEE, K. (1983). Uniformity of energy deposition for laser driven fusion. *J. Appl. Phys.* **54**, 3662-3671.

SKUPSKY, S., MAROZAS, J.A., CRAXTON, R.S., BETTI, R., COLLINS, T.J.B., DELETREZ, J.A., GONCAROV, V.N., McKENTY, P.W., RADHA, P.B., KNAUER, J.P., MARSHALL, F.J., HARDING, D.R., KILKENNY, J.D., MEYERHOFER, D.D., SANGSTER, T.C. & McCRORY, R.L. (2004). Polar direct drive on the National Ignition Facility. *Plasma Phys.* **11**, 2763.

TEMPORAL M. & CANAUD, B. (2009). Numerical analysis of the irradiation uniformity of a directly driven inertial confinement fusion capsule. *Eur. Phys. J. D* **55**, 139.

TEMPORAL, M., CANAUD, B. & LE GARREC, B.J. (2010a). Irradiation uniformity and zooming performances for a capsule directly driven by a $32 \times 9$ laser beams configuration. *Phys. Plasmas* **17**, 022701.

TEMPORAL, M., CANAUD, B., LAFFITE, S., LE GARREC, B.J. & MURAKAMI, M. (2010b). Illumination uniformity of a capsule directly driven by a laser facility with 32 or 48 directions of irradiation. *Phys. Plasmas* **17**, 064504.

TEMPORAL, M., RAMIS, R., CANAUD, B., BRANDON, V., LAFFITE, S. & LE GARREC, B.J. (2011a). Irradiation uniformity of directly driven inertial confinement fusion targets in the context of the shock-ignition scheme. *Plasma Phys. Contr. Fusion* **53**, 124008.

TEMPORAL M. & CANAUD, B. (2011b). Stochastic homogenization of the laser intensity to improve the irradiation uniformity of capsules directly driven by thousands laser beams. *Eur. Phys. J. D* **65**, 447.

TEMPORAL, M., CANAUD, B., GARBETT, W.J., PHILIPPE, F. & RAMIS, R. (2013). Polar direct drive illumination uniformity provided by the Orion facility. *Eur. Phys. J. D* **67**, 205.

TEMPORAL, M., CANAUD, B., GARBETT, W.J. & RAMIS, R. (2014a). Numerical analysis of the direct drive illumination uniformity for the Laser MegaJoule facility. *Phys. Plasmas* **21**, 012710.

TEMPORAL, M., CANAUD, B., GARBETT, W.J. & RAMIS, R. (2014b). Irradiation uniformity at the Laser MegaJoule facility in the context of the shock ignition scheme. *High Power Laser Science and Engineering* **2**, e8.
M.Temporal et al.　　　　　　　　　　28 June 2014　　　　　　　　　　8/8